# Architecture for Integrating Learning Platforms Using Adapter

Ali B. Dauda[1*], Abubakar A. Idris[2], Abdulaziz. I. Ibrahim[1], Suleman Umar[1], Peter Yohanna Mshelia[3]

[1]University of Maiduguri, ICT Center, Nigeria
[2]Federal College of Agricultural Produce Technology, Kano, Nigeria
[3]Nigeria Army University Biu, Nigeria

ali.dauda@unimaid.edu.ng, auwalabubakaridris@gmail.com, peter@unimaid.edu.ng, abdulaziz@unimaid.edu.ng
yasule@unimaid.edu.ng

*Abstract*— The advantage of the electronic and mobile learning platforms is the dissemination of learning contents with ease. But these platforms operate differently to exchange the learning contents from the server (educator's site) to the clients (learner's site). Integrating these learning platforms to operate as a single platform and exchange the contents based on learners' request could improve the learning efficiency and reduce the operational cost. This work introduces a Web services approach based on client-server model to develop an integrated architecture that join the two learning platforms. In this paper, the architecture of the learning platforms is presented and explained. Furthermore, an adapter in a form of web service is develop as a fuse between the server and the client. Finally, the process of using the web services to unify the two learning architectures using the adapter is demonstrated and explained.

## I. Introduction

The Internet has given rise to communication regardless of location and time. Among the benefits is learning via the Internet. Internet-based learning has provided paybacks for both the learners and educators by breaching barriers and cutting cost. It brakes the impediment of learning process by increasing the level of teacher-to-student interaction and making learning resources available anywhere and anytime. Mainly, two process of learning dominated the Internet learning sphere: the electronic learning and mobile learning.

The learning process of using computer to share information is referred to as electronic learning or e-learning. This encompasses learning via Computer, TV, Video, Radio, CD/DVDs, Mobile phones, Tablets and other electronic learning devices [1]. Learning over the Internet is one aspect of e-learning that deals with learning using the network devices to communicate and deliver learning contents.

Another aspect is the learning via mobile devices is referred to as mobile learning or m-learning. M-learning provides mode that supports process of learning through mobile devices in a just-in-time [2]. Mobile devices include such as smart phones, tablets and iPad.

There are numerous e-learning and m-learning services across the world, some coupled together and others work independently as applications to deliver the learning content [3]. Having an architecture that incorporates both learning platforms will offer an understanding of the major requirement for establishing a standard integrated learning environment. An enhanced integrated architecture will improve the performance, maintainability and security of development of the learning system.

This paper proposed a Web services architecture that unifies the el-learning and m-learning platforms as a single service. The paper provides an architecture model of a generic architecture using an adapter to integrate the electronic and mobile learning process.

The rest of the paper is organized as follows; Section 2 provided related work on e-learning and m-learning architectures. Section 3 presented and described the e-learning architecture and the m-learning architecture. Section 4 presents the integration of the Learning platforms with Web services, while Section 5 concluded the paper is concluded the paper.

## II. Related Work

Research have been conducted on unifying e-learning and m-learning to operate as an entity. There are several studies on architectures, frameworks and models to address the unification of the learning services or validate the need for digital learning.

Dagger, O'Connor, Lawless, Walsh, and Wade [4] proposed a framework on Service-oriented e-learning by providing a model of addressing e-learning in Service-oriented approach. The aim of the research is to achieve interoperability in the e-learning platforms. The authors provided a framework by using Web services using SOAP messaging with HTTP and XML to provide seamless and interoperable platform for online based learning. The authors concluded by suggesting that a proper Service-oriented can enhance the future e-learning.

Study by [5] discussed the m-learning as a learning content delivery but can only be effective in three ways; pure connection, pure mobility and the hybrid of the two pures. They further discussed the use of WAP and GPRS to access the learning content. Their work was based purely on the non-interactive output of the learning content using web services. The study provides an architectural view of e-learning architecture and how web services can be used in the e-learning environment to disseminate transfer content to the learners' endpoint.

A related study by [6] suggested the integration of m-learning on e-learning environment. They argued that mobile devices can be use like a classroom setup. The study emphasized on provision of real-time interaction between teacher and student using mobile device. The study provided an adaptive learning process of integrating the mobile learning system into conventional learning process. The study worked on the integration of mobile learning content to work as part of the e-learning processes. Although, the architectural view of the process was presented but not discussed.

M-learning architecture was proposed by [7]. The architecture comprised of Web services on top of learning management system. The Web services takes the responsibility of transporting the learning content with mobile specific functionalities. The functionalities comprised of context discovery, mobile content management and adaptation, and packaging and synchronization. These features gave the architecture a form of generic architecture for m-learning for learning management system. The outcome is an architecture that automatic selects services and render content relevant to what the user device can accommodate.

Although studies have been conducted to integrate the learning platforms, the architecture for the integration is not fully addressed. This paper presents the architecture for integration of the learning platforms using adapter. The adapter employed web services standard for the service connections between the learner and educator endpoints.

### III. LEARNING INTEGRATION ARCHITECTURE

This section presents the Web services overview and the architecture for the individual learning platforms. It demonstrates the content of the platforms and how each is accessed to deliver the content via web services server/client communication.

*A. Web services architecture overview*

Web services are message-based set of operations that are accessible on a network via standardized formal XML. This a framework that hosts emerging standards like HTTP, XML, SOAP, WSDL and UDDI. One of the major advantages of web services is its ability of integrating web applications more easily, fast and inexpensive. Integration technologies like DCOM and CORBA provide a standard for integrating well tightly coupled applications at high levels as their major limitation [8], in contrast, Web services provides loosely coupled components [9]. Figure 1.0 shows the components of Web services based on software-oriented architecture.

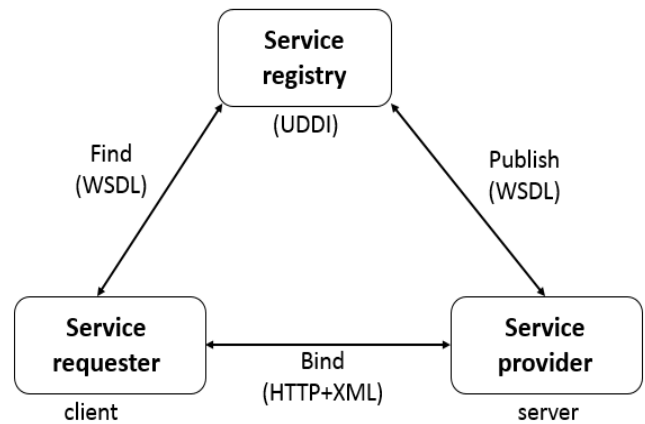

Figure 1.0 Web services Architecture (Newcomer and Lomow, 2005)

The service request is established by the requester and the provider published the service and post it to the service registry as WSDL file. The service is discovered by WSDL and generates a Web service client. The requester finds the WSDL file through the registry UDDI and consume the service. Once the service is discovered by the requester, the client interacts with the server based on the WSDL description to bind and consume the service. As accorded by [10], the services are set of information and operations defined in form of XML-based SOAP messages.

In Figure 1.0, HTTP is the transfer protocol, XML is the messaging protocol. The WSDL is the actual standard for XML-based service description while the UDDI is the standard XML-based service directory [11].

The basic principle is that Web services has provider – consumer model which implement service by publicising in a directory – the UDDI. This service is registered at the UDDI directory that serves as all-purpose directory. Service user search for the service or multi-services that suits the particular need. The WSDL on the other hand, describes how the service will be consumed by the Web service and produce proxy to communicate with Web services through SOAP [12].

*B. E-Learning web service architecture*

Share-ability in both synchronous and asynchronous mode of learning content is the major aim of e-learning. Content can vary in size and usage, for instance audio, visual and streaming data or user interaction such as editing and sharing of document from whiteboard. The e-learning architecture presents the transfer of e-learning web content from the producer (server) to the consumer (client) end through the interaction of web service [13]. Figure 2.0 shows the e-learning architecture and how the contents are produced and exchanged in form of web service request/response. As seen in the figure the login, user and authoring services find a related system registry and published request through either delivery, content or search services to the client using a service broker. Sharing of information can be unicast or multicast style depending on the type of audience and content request.

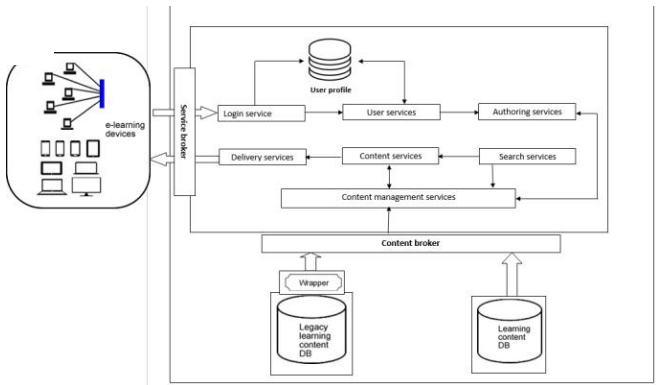

Figure 2.0 e-learning web services architecture

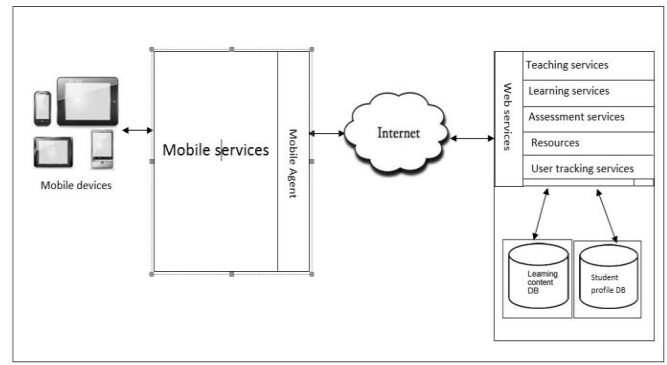

Figure 3.0 m-learning architecture

As shown figure 2.0, the e-learning devices requests for content from the server via the services broker. The broker pass the request to the login service for user authentication. Verified login are authorised and request is then pass to the content management services (CMS). The CMS fetch the appropriate learning content from the database through content broker and exchange the content via the delivery services.

### C. M-Learning web services architecture

A new paradigm in this learning transformation is the mobile learning or m-learning [14]. M-learning is expanding in the field of education and training due to its coverage, accessibility, cost effectiveness and just-in-time result particularly in collaborative set-up.

Generally, its definition involves the use of any communication device that allow synchronous or asynchronous communication. This become important due to explosion of high processing cheap GPRS enabled mobile phones. Users can access learning content irrespective of location and time [15]. Though, there are numerous works on architecture covering m-learning as a single platform. Mobile learning content are deployed using m-learning service architecture due to the nature of the client's devices miniature size, input style and screen resolutions. Figure 3.0 shows the architecture of a service based of m-learning.

Client's device requests this service and use and terminate the service at its own time phase. The advantage here is, despite the device-to-device operability, it can be executed on different platforms and devices.

According to [16], the predominant use of web services in the deployment of learning content is clear and has gained ground. But the due to miniature size input style, memory capacity and screen resolutions of these mobile devices, not all contents can be delivered on these devices. As such, contents are delivered limited to the capability of mobile devices. Web services access the learning content on the server, filter and send the service response based on the client request. The services are provided as HTML wrapped XML-messages.

Figure 3.0 shows how the m-learning content are exchanged to the user. The mobile devices request for content using the device mobile services on the client side. The mobile agent in the mobile services use the request parameters and request for appropriate content through the HTTP. The Web services in the server side assess the request and appropriately match it with learning services. The Web services then and fetch the learning content and push it through via the HTTP. The response for the request is a learning content for the requesting client device.

### IV. LEARNING SYSTEM INTEGRATION WITH WEB SERVICES

Combining the electronic and mobile learning paradigms is essentials in order to provide effective learning process. To achieve these, there is need for a unified architecture to incorporate the requirements for both learning contents. This is achievable by combining the contents and modify the request based on the device type and content request. The unified architecture could have the capable to detect appropriate client device type and deploy the right content to the device with almost the same content with other learning devices. The unification can be achieved by integrating the content through a middle generic application in the form adaptor.

### A. The Proposed Integration adapter

The adaptor functions as a translating mechanism between the devices and translates the learning content from the server to a formatted content readable on the client device.

Web services operates based on service-oriented architecture which uses SOAP and XML. The SOAP in the web services is the messaging protocol that make a remote procedure (RPC) calls from the server and the devices by formatting and posting XML messages via the HTTP as a request. The SOAP-RPC invokes a function from client device to behave like a local function on the server. The client calls the web service by sending a set of parameters and receiving return values as services [17].

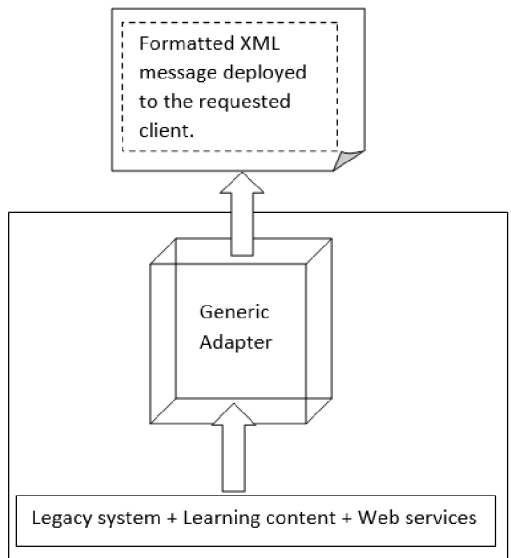

Figure 4.0 Integration adapter

The adaptor functions as a proxy for the requested web service. The server may send multiple response to many clients as possible, in multiple forms. For example, a mobile phone requesting a data from excel and a laptop requesting a web form to access a course forum. The Adaptor detects the device and send an XML message appropriate to the machine/device via a HTTP.

The e-learning content and the m-learning content exist on the same architectural platform. The client makes a request through the web services without going through the adapter. The response to the client request then went through the adapter set-up which access the content, package and deliver the response in an XML+HTML style. The XML describes the representation of the content and the HTML presents the output in a readable manner. The line of codes in Table 1.0 show a simple implementation of the adapter XML tagged into HTML rendering a formatted list of students on DCS 202 forum. The final output is a list of students their matric numbers.

Table 1.0: Server response as XML wrapped Messages in HTML

```
<html>
<body>
<script>
if (window.XMLHttpRequest)
  {// code for IE7+, Firefox, Chrome, Opera, Safari
  xmlhttp=new XMLHttpRequest();  } else
  { // code for IE6, IE5
  xmlhttp=new ActiveXObject("Microsoft.XMLHTTP");  }
xmlhttp.open("GET","course_forum.xml",false);
xmlhttp.send();
xmlDoc=xmlhttp.responseXML;
document.write("<table border='1'>");
var x=xmlDoc.getStudents_listByName("DCS 202 ");
for (i=0;i<x.length;i++)
{   document.write("<tr><td>");
  document.write(x[i]. getStudents_listByName ("MATRIC-NO")[0].childNodes[0].nodeValue);
  document.write("</td><td>");
  document.write(x[i]. getStudents_listByName ("NAME")[0].childNodes[0].nodeValue);
  document.write("</td></tr>");  }
document.write("</table>");
</script>
</body>
</html>
```

## V. CONCLUSION

The unification of a single architecture for both e-learning and m-learning provides a cost effective and time efficient system. Web services connects heterogeneous systems regardless of platforms. The SOAP in the Web services exchanged the content in form of XML message regardless of the receivers' environment. The learning content is rendered to be readable by the device receiving the content. The adapter acts as a bridge that process the content of the learning platforms based on the request from a client device or machine. The adapter functions as a remote procedure call invoked by the client that triggered the server to produce an XML messages as services. The XML messages are wrapped into an HTML tags and exchanged to the client that requested the service.